 \documentclass[final,5p,times,twocolumn, number]{elsarticle}
 \biboptions{compress}

\usepackage{amssymb}
\usepackage{amsmath}
\usepackage{amsfonts}
\usepackage{mathtools}
\usepackage{physics}

\newcommand{\ms}{\overline{\textrm{MS}}}
\newcommand{\cw}{{\rm{CW}}}
\newcommand{\Vtree}{V^{(0)}}
\newcommand{\Vone}{V^{(1)}}
\newcommand{\f}{\varphi}

\newcommand{\fex}{\langle\f\rangle}
\newcommand{\lms}{\lambda_{\ms}}
\newcommand{\los}{\lambda_{\mathrm{OS}}}
\newcommand{\mms}{m^2_{\ms}}
\newcommand{\mos}{m^2_{\mathrm{OS}}}
\newcommand{\rcw}{r_{\cw}}
\newcommand{\vms}{v_{\ms}}
\newcommand{\vcw}{v_{\cw}}
\newcommand{\lcw}{\lambda_{\rm{CW}}}
\newcommand{\mcw}{m^2_{\rm{CW}}}

\journal{Physics Letters B}
\usepackage{hyperref}

\begin{document}

\begin{frontmatter}

\title{Radiative symmetry breaking from the on-shell perspective}

\author[1]{Bartosz S\'{o}jka}
\author[1]{Bogumi{\l}a \'{S}wie\.{z}ewska}
\affiliation[1]{organization={Faculty of Physics, University of Warsaw},
            addressline={Pasteura 5}, 
            city={Warsaw},
            postcode={02-093}, 
            country={Poland}}

\begin{abstract}
Models with classical scale symmetry, which feature radiative symmetry breaking, generically lead to a supercooled first-order phase transition in the early Universe resulting in a strong gravitational-wave signal, potentially observable by LISA. This is thanks to the absence of mass terms in the potential and the resulting logarithmic structure of the zero-temperature effective potential. It~is known that imposing a symmetry at classical level does not prohibit breaking it by quantum corrections. In the case of scale symmetry, a mass term can in principle appear through renormalisation. This is not the case in the commonly used $\overline{\mathrm{MS}}$ or Coleman--Weinberg schemes. In this work, we renormalise a scale-invariant model in the on-shell scheme to check whether parameterising it with the physical masses will introduce mass terms to the potential. We find that indeed mass terms appear for an  arbitrary choice of the physical masses. However, we formulate an on-shell condition for radiative symmetry breaking, sufficient and necessary for the cancellation of mass terms in the renormalised potential, yielding a~logarithmic potential needed for supercooled phase transitions.
\end{abstract}


\end{frontmatter}

\section{Introduction}
\label{introduction}

Cosmological phase transitions (PT) are among the early-Universe phenomena that will likely be probed via gravitational waves (GW), in particular by LISA~\cite{LISACosmologyWorkingGroup:2022jok,Caprini:2015zlo,Caprini:2019egz}. The electroweak symmetry breaking proceeds via a crossover within the Standard Model (SM)~\cite{Kajantie:1996mn, Gurtler:1997hr,Csikor:1998eu}, therefore no observable GW signature is expected in this scenario. However, first-order PTs can be realised in extensions of the SM, which are needed to solve the long-standing puzzles of dark matter, baryon asymmetry of the Universe, etc. Numerous models of beyond-SM (BSM) physics are subject to detailed scrutiny to determine the nature of the predicted PT and the associated GW signal. 
Such a signal would be a clear sign of BSM physics and a source of information about fundamental interactions, complementary to collider searches. Even a negative outcome, with no PT-like signal, would give us important information, constraining or excluding certain BSM models.

A supercooled PT is a scenario well-suited for being tested with LISA. It proceeds at temperatures much below the critical temperature, leading to a substantial energy release and a strong GW signal. This kind of transition is accommodated by models with nearly conformal dynamics or classical scale symmetry~\cite{Randall:2006, Konstandin:2010, Konstandin:2011,Hambye:2013,Jaeckel:2016}. Results for various models with classical scale symmetry indicate that GW from a supercooled PT will be well within experimental reach~\cite{Hambye:2013, Jaeckel:2016, Jinno:2016, Marzola:2017, Hashino:2018, Baldes:2018, Prokopec:2018,Marzo:2018, 
Mohamadnejad:2019vzg, Kang:2020jeg, Ellis:2020, Azatov:2020nbe, Mohamadnejad:2021tke, Dasgupta:2022isg, 
Kierkla:2022odc, Frandsen:2022klh, Kierkla:2023von, Chun:2023ezg}. A particular example is a classically scale invariant model with a new SU(2) gauge group for which the whole allowed parameter space yields a GW signal with a significant signal-to-noise ratio~\cite{Kierkla:2022odc,Kierkla:2023von}. Thus, in case of non-observation of a PT-like signal, the whole model can be falsified. Moreover, a supercooled PT has been one of the suggested explanations~\cite{ Ratzinger:2020koh, NANOGrav:2021flc, Gouttenoire:2023bqy, Madge:2023dxc, Figueroa:2023zhu, Ellis:2023oxs} of the stochastic GW background recently observed by the pulsar timing array collaborations~\cite{NANOGrav:2023gor, EPTA:2023fyk,Reardon:2023gzh}. However, it is not easy to construct a concrete model that would fit the data and would not contradict other observations~\cite{Bai:2021ibt, Bringmann:2023opz}.

The characteristic feature that makes models with classical scale invariance particularly suited for strong first-order PTs is a logarithmic potential. The lack of a term quadratic in the field (that would be associated with a mass term in the potential), makes the potential very flat around the origin as it is dominated by the term quartic in the field. In such a setting, even small thermal corrections introduce a barrier that delays the tunnelling from the symmetric to the broken phase until very low temperatures. This causes supercooling and results in a large energy release and a strong GW signal. It is known, however, that if a classical symmetry is present in a theory, nothing prevents it from being broken by loop corrections. Indeed, within classically conformal models, masses of particles are generated via radiative symmetry breaking (RSB). This can be realised without an explicit mass term in the potential. But there is no reason, in principle, why such a term would not arise at one-loop. If it does, it could entirely change the physical picture of a strong, supercooled PT.

In the widely used modified minimal subtraction ($\ms$) renormalisation scheme and in the scheme introduced in the seminal paper studying RSB by Coleman and Weinberg~\cite{Coleman:1973jx}, a mass term does not appear at one-loop level. But, as we explain in detail in the main body of the text, these schemes are tailored to preserve the scaling symmetry at the level of the Lagrangian. However, in the physical world scales exist and in particular, masses exist. Then a question arises -- does parameterising the model with physical masses require the inclusion of a mass counterterm that would explicitly break scale invariance? The aim of the present work is to renormalise a classically scale-invariant model in the on-shell (OS) scheme and study the consequences of such parameterisation for the PT and the resulting GW. Moreover, this analysis will allow us to evaluate how well the commonly used running masses (computed within the $\ms$ scheme) approximate the physical masses. This may seem to be a trivial question that should have been long answered given the number of works on models with classical scale symmetry. However, to our knowledge, it has not been so far addressed in the literature. Motivated by the important phenomenological consequences of supercooled PTs we aim to fill this gap in the present work.

In this article, we will analyse in detail the renormalisation of massless scalar quantum electrodynamics (QED) as a minimal toy model allowing for RSB~\cite{Coleman:1973jx}. The model is introduced in section~\ref{sec:model}. Further, we will review different renormalisation schemes in section~\ref{sec:schemes}. Our approach to the on-shell renormalisation of scalar QED is presented in section~\ref{sec:OS}. Next, we relate the obtained results to the more familiar $\ms$-results in section~\ref{sec:MS-OS} and discuss the relation between loop-generated and tree-level mass terms in section~\ref{sec:mass-terms}. Finally, we summarise our findings and discuss their implications for GWs from supercooled PTs in section~\ref{sec:summary}.

\section{Massless scalar quantum electrodynamics\label{sec:model}}

Starting from the seminal paper of Coleman and Weinberg~\cite{Coleman:1973jx}, massless scalar quantum electrodynamics (QED), also known as the massless Abelian Higgs model, has been an archetypal model for the mechanism of RSB (also known as dimensional transmutation). It consists of a complex scalar charged under a U(1) gauge group and a gauge field -- a photon -- originating thereof. The Lagrangian is given by:
\begin{equation}
    \mathcal{L}= -\frac{1}{4}F_{\mu\nu}F^{\mu\nu}+\left(D_\mu\phi\right)^* D^\mu\phi-\Vtree,\label{eq:QED-Lagrangian}
\end{equation}
where $F_{\mu\nu}$ is the usual energy-momentum tensor for the gauge field, the covariant derivative is given by $D_{\mu}=\partial_{\mu}+ieA_{\mu}$, $e$ is the U(1) charge and $\phi$ denotes the complex scalar field. The tree-level potential is given by
\begin{equation}
    \Vtree(\phi)=\lambda(\phi^*\phi)^2.
\end{equation}
We can use the global U(1) symmetry to rotate the complex field such that the vacuum expectation value (VEV) is located along the real direction and write the classical contribution to the effective potential as
\begin{equation}
     \Vtree(\f)=\frac{1}{4}\lambda\f^4,\label{eq:V-tree}
\end{equation}
where $\f$ denotes the classical (background) field. The Abelian Higgs model possesses classical scale (conformal) symmetry as no dimensionful parameters are present at tree level. Classically, the model features only a symmetric minimum at $\f=0$.

To study the vacuum structure beyond tree level, one needs to consider the effective potential. While the scalar contributions to the effective potential are negligible for perturbative couplings~\cite{Coleman:1973jx}, including loop corrections from the gauge boson modifies the structure of the potential. The one-loop correction to the effective potential reads (in the Landau gauge and using dimensional regularisation in $D=4-\varepsilon$ dimensions)
\begin{equation}
    \Vone(\f)=\kappa e^4 \f^4\left(\log\frac{e^2\f^2}{\mu^2}-\frac{5}{6}+\eta\right),\label{eq:V-one-loop-correction}
\end{equation}
where we introduced a short-hand notation for the loop factor $\kappa=\frac{3}{64 \pi^2}$, the term $e\f$ corresponds to the field-dependent mass of the photon and the infinite part is contained in $\eta=-\frac{2}{\varepsilon}+\gamma_E-\log(4\pi)$, with $\gamma_E=0.5772$ being the Euler constant. One should note, that the one-loop term computed using dimensional regularisation does not contain infinities proportional to $\f^2$. This would not be the case if cutoff regularisation was used~\cite{Coleman:1973jx}. To make the potential finite, we need to split the bare couplings into the sum of renormalised ones and the counterterms that will cancel infinities as
\begin{equation}
    \lambda_{\mathrm{bare}}=\lambda +\delta\lambda, \quad m^2_{\mathrm{bare}}=m^2+\delta m^2.
\end{equation}
Both the renormalised couplings and the counterterms depend on the renormalisation scheme. Even if we assume that the renormalised mass term vanishes, the dimensionful mass counterterm  $\frac{1}{2}\delta m^2 \f^2$ has to be included since the scale symmetry is only imposed on the model at classical level. Since there is no infinity proportional to $\f^2$, $\delta m^2$ can serve to perform finite renormalisation of the mass. In principle, also the field strength has to be renormalised by performing the shift $\f^2\to Z \f^2$. However, the effective potential is the zero-momentum part of the effective action so we can assume that $Z=1$ at the renormalisation scale at work. Also note that we do not need to renormalise $e$, as it only appears in the effective potential at one-loop level.

The full renormalised one-loop potential reads
\begin{equation}
    V=\Vtree+\Vone+V_{\mathrm{CT}},\quad V_{\mathrm{CT}}(\f)=\frac{1}{2}\delta m^2 \f^2 + \frac{1}{4}\delta \lambda\f^4.\label{eq:V-eff}
\end{equation}
The values of the counterterms depend on the renormalisation scheme. In the next section, we review the popular choices for the renormalisation schemes.

After introducing the effective potential in the form that will be used throughout this article, for the sake of generality we will now comment on the robustness of the obtained predictions with respect to the change of the gauge. To address this issue, one can consider the effective potential computed in the general Fermi gauge with the gauge fixing term given by
\begin{equation}
    \mathcal{L}_{\rm gf}=-\frac{1}{2\xi}\left(\partial_{\mu} A^{\mu}\right)^2.
\end{equation}
An advantage of this choice of gauge fixing is that it preserves the U(1) and scaling symmetry of the original Lagrangian. With this choice, the explicit gauge-dependence of the effective potential becomes apparent in the contributions from the mixed gauge-Goldstone bosons~\cite{Andreassen:2014eha,Espinosa:2016uaw} (for a pedagogical derivation see ref.~\cite{Andreassen:2013hpa}, also see ref.~\cite{Loebbert:2018xsd} for results in a similar model),
\begin{equation}
    V_{\xi}(\f)=\frac{1}{64\pi^2}\sum_{+,-}M_{\pm}(\f)^4\left(\log\frac{M_{\pm}(\f)^2}{\mu^2}-\frac{3}{2}\right),
\end{equation}
where $M_{\pm}(\f)^2=\frac{1}{2}\left(M_G(\f)^2\pm \sqrt{M_G(\f)^2\left(M_G(\f)^2-4\xi e \f^2\right)}\right)$, with $M_G(\f)^2=3\lambda \f^2$. In the Landau gauge, which corresponds to $\xi=0$, this reduces to $M_+=M_G$, $M_-=0$. If we now anticipate the scaling $\lambda\sim e^4$ (see eq.~\eqref{eq:lambda-e-MS-fixed-scale}) typical for RSB, we see that the terms depending on $\xi$ scale as $e^6$ (and the $\xi$-independent parts of $V_{\xi}$ scale as $e^8$). Therefore, working at the leading order of $\mathcal{O}(e^4)$, we can neglect these terms. This yields the effective potential of eq.~\eqref{eq:V-eff}, which is gauge-independent~\cite{Andreassen:2014eha,Espinosa:2016uaw}. Therefore, all the results presented in this work do not depend on the choice of the gauge. The $\xi$-dependence at $\mathcal{O}(e^6)$ must be included together with two-loop contributions of that order but as was shown in ref.~\cite{Kang:1974yj} (in the Coleman-Weinberg renormalisation scheme) in the mass ratio the $\xi$ dependence cancels out.

\section{Different renormalisation schemes\label{sec:schemes}}

The infinities present in loop computations can be removed by imposing renormalisation conditions. Different sets of conditions can be chosen, leading to diverse renormalisation schemes. The most popular one, due to its convenience, is the modified minimal subtraction scheme ($\ms$). In the context of RSB, the original scheme introduced by Coleman and Weinberg~\cite{Coleman:1973jx} is also used sometimes. The parameters defined within these schemes cannot, however, be considered physical. The scheme that relates the parameters of the potential to the on-shell masses of physical particles is the so-called on-shell (OS) scheme. In this section, we review these renormalisation schemes.

\paragraph{Modified minimal subtraction scheme\label{sec:MS}} Within the $\ms$ scheme the counterterms are required to simply cancel infinities (plus some constant terms) proportional to respective powers of $\f$. Since no divergent terms proportional to $\f^2$ are generated at one-loop level  when dimensional regularisation is used (see eq.~\eqref{eq:V-one-loop-correction}), the $\delta m^2$ counterterm is equal to zero. We have
\begin{equation}
    \begin{dcases}
        \delta \mms=0,\\
        \delta \lms = -4\kappa e^4\eta,
    \end{dcases}\label{eq:MS-counterterms}
\end{equation}
and the one-loop renormalised potential takes the following form
\begin{equation}
    V(\f)=\frac{1}{4}\lms\f^4+\kappa e^4 \f^4\left(\log\frac{e^2\f^2}{\mu^2}-\frac{5}{6}\right).\label{eq:MS-potential}
\end{equation}
We introduced a subscript on the coupling which indicates the scheme in which it is defined.

We can now study the symmetry breaking. The minimisation condition, with $\fex=\vms$ yields two solutions
\begin{equation}
    \lms=-4 \kappa e^4 \left(\log\frac{e^2\vms^2}{\mu^2}-\frac{1}{3}\right)\label{eq:lambda-e-MS}
\end{equation}
or $\vms=0$. The special property of models with classical scale invariance is that from the minimisation of the potential we do not obtain the value of the VEV, but rather a relation between the couplings. Setting $\mu=e\vms$, it further simplifies to
\begin{equation}
     \lms=\frac{4}{3}\kappa e^4=\frac{e^4}{16 \pi^2}\label{eq:lambda-e-MS-fixed-scale}.
\end{equation}
This relation is at the heart of understanding RSB -- the one-loop correction can modify the potential quantitatively if the tree-level (parameterised by $\lambda$) and the one-loop (parameterised by $e^4$) contributions belong to the same order in perturbation theory. In this case, one should not expand in loops, but rather in powers of couplings, using the scaling of eq.~\eqref{eq:lambda-e-MS-fixed-scale}. This is important also to preserve gauge independence of the effective potential at the minimum in perturbative computations beyond leading order, see refs.~\cite{Andreassen:2014eha,Espinosa:2016uaw}.

Using the relation of eq.~\eqref{eq:lambda-e-MS} we can simplify the $\ms$-renormalised effective potential of eq.~\eqref{eq:MS-potential} to the following form
\begin{equation}
   V(\f)= \kappa e^4 \f^4 \left(\log\frac{\f^2}{\vms^2}-\frac{1}{2}\right).
\end{equation}
Note that formally we have not set the scale to $e\vms$. The scale dependence in $\lms$\; in eq.~\eqref{eq:lambda-e-MS} cancelled the scale dependence in the effective potential. The resulting potential is parameterised using just two quantities: $e$ and $\vms$ -- the presence of a dimensionful parameter signals dimensional transmutation. 

Using the effective potential we can approximate the mass of the scalar by the second derivative of the potential evaluated at the minimum (the running mass, denoted with a bar as $\overline{M}_S^2$),
\begin{equation}
    \overline{M}_S^2=V''(\vms)=8\kappa e^4 \vms^2.
\end{equation}
Now, knowing the mass of the vector, $\overline{M}_V=e\vms$, we can compute the ratio of masses
\begin{equation}
    \frac{\overline{M}_S^2}{\overline{M}_V^2}=8 \kappa e^2=\frac{3e^2}{8\pi^2}\equiv \rcw. \label{eq:mass-ratio}
\end{equation}
This ratio of masses is said to be a prediction of the massless scalar QED~\cite{Coleman:1973jx}. We will check whether it holds for the physical masses (not just the running ones).

\paragraph{Coleman-Weinberg (CW) scheme} The CW scheme~\cite{Coleman:1973jx} consists of the following conditions
\begin{equation}
\begin{dcases}
     \left.\frac{\dd^2 V}{\dd\f^2}\right|_{\f=0}=0,\\*
    \left.\frac{\dd^4 V}{\dd\f^4}\right|_{\f=M}=6\lcw.\label{eq:CW-conditions}
\end{dcases}
\end{equation}
One should note that the first condition explicitly enforces that the renormalised mass term vanishes. The other condition defines the coupling at a certain energy scale $M$. One cannot define it at $\f=0$ in analogy to the first one, since the logarithmic term is ill-defined there. The counterterms read\footnote{In ref.~\cite{Coleman:1973jx} the cutoff regularisation was used, thus a quadratic infinity arose which had to be cancelled by a non-zero mass counterterm. Here we use dimensional regularisation, thus $\delta \mcw=0$.}
\begin{equation}
\begin{dcases}
     \delta \mcw=0, \\
   \delta \lcw = -4 \kappa e^4 \left(\log\frac{e^2 M^2}{\mu^2}+ \frac{10}{3}+\eta\right).
\end{dcases}\label{eq:CW-counterterms}
\end{equation}
The potential, after expressing $\lcw$ in terms of the coupling $e$ with the use of the minimisation condition has the same form
as in the $\ms$ case, eq.~\eqref{eq:MS-potential}, with $\vms$ replaced by $\vcw$. Moreover, the predicted ratio of masses is the same as in $\ms$, hence it is denoted by $\rcw$ in  eq.~\eqref{eq:mass-ratio}.


\paragraph{On-shell scheme} The on-shell scheme is considered to be the physical renormalisation scheme as it requires the renormalised mass to correspond to the pole of the propagator, with residue equal to one. A~condition requiring that the VEV of the scalar field corresponds to the physical one is often added. The general renormalisation conditions can be summarised as:
\begin{equation}
    \begin{dcases}
    \left.\frac{\dd V}{\dd\f}\right|_{\f=v}=0,\\[4pt]
    G^{(2)}(p^2=M^2)=0,\\[4pt]
    \left.\frac{\dd G^{(2)}}{\dd p^2}\right|_{p^2=M^2}=1,\label{eq:on-shell-general}
    \end{dcases}
\end{equation}
where $G^{(2)}(p^2)$ is the two-point function (the inverse of the propagator) and $M$ corresponds to the physical mass. Most commonly, in the case of one-loop renormalisation, $M$ and $v$ are identified with the tree-level parameters. In this case, the conditions of eq.~\eqref{eq:on-shell-general} can be further simplified. Using the expression for the loop-corrected two-point function
\begin{equation}
    G^{(2)}(p^2)=p^2-M^2-\Sigma(p^2),
\end{equation}
with $\Sigma(p^2)$ being the one-loop self-energy and assuming that the tree-level mass corresponds to the physical mass $M$ and that the physical VEV is the tree-level one, we obtain
\begin{equation}
    \begin{dcases}
    \left.\frac{\dd\Vone}{\dd\f}\right|_{\f=v}=0,\\[4pt]
    \Sigma(p^2=M^2)=0,\\[4pt]
    \left.\frac{\dd\Sigma}{\dd p^2}\right|_{p^2=M^2}=0.\label{eq:on-shell-sigma}
    \end{dcases}
\end{equation}
If one is interested in the effective potential, one can set the wave-function renormalisation constant $Z$ to one, at some given renormalisation scale $\mu$ and neglect the last condition. Furthermore, the second derivative of the one-loop correction to the effective potential corresponds to the zero-momentum part of the self-energy, thus
\begin{equation}
    \Sigma(p^2)=\Sigma(0)+\Sigma(p^2)-\Sigma(0)=\left.\frac{\dd^2\Vone}{\dd \f^2}\right|_{\f=v}+\Delta \Sigma(p^2).
\end{equation}
If $\Delta \Sigma(M^2)$ can be considered negligible, the set of renormalisation conditions can be rewritten as
\begin{equation}
    \begin{dcases}
    \left.\frac{\dd\Vone}{\dd\f}\right|_{\f=v}=0,\\[4pt]
    \left.\frac{\dd^2\Vone}{\dd\f^2}\right|_{\f=v}=0.
    \label{eq:on-shell-Vtree}
    \end{dcases}
\end{equation}
This is the form of the on-shell renormalisation conditions commonly used in computations making use of the effective potential. We cannot, though, apply these conditions in the case of classically scale-invariant models since there is no tree-level VEV or mass to refer to. We devote the next section to studying the OS renormalisation of the massless scalar QED.

\section{On-shell renormalisation\label{sec:OS}}
In the absence of tree-level symmetry breaking, we will introduce renormalisation conditions which refer to the full one-loop effective potential, not just the one-loop correction as discussed in section~\ref{sec:schemes}. In~the former approach, one could use the tree-level relations  given by the minimisation condition and the second derivative of the potential to determine the VEV and the masses using Lagrangian parameters. Here, the one-loop mass and VEV will be used to fix the counterterms so we will not be able to express the physical masses in terms of the Lagrangian parameters. Nonetheless, we will be able to obtain the full one-loop effective potential parameterised by the physical masses. As the input parameters, we choose the mass of the vector $M_V$, the mass of the scalar $M_S$ and the electric charge $e$. One should note that we parameterise the model with 3 free parameters, not 2 as was in the case of $\ms$ and CW schemes.

The OS renormalisation conditions read
\begin{equation}
    \begin{dcases}
    \left.\frac{\dd V}{\dd\f}\right|_{\f=M_V/e}=0,\\[4pt]
    \left.\frac{\dd^2 V}{\dd\f^2}\right|_{\f=M_V/e}=M_S^2.
    \label{eq:on-shell-Veff}
    \end{dcases}
\end{equation}
As already explained in the previous section, they state that $M_V/e$ corresponds to the VEV of the scalar field and $M_S$ corresponds to the pole of the scalar propagator in the zero-momentum limit. In the OS scheme, we obtain the following relations
\begin{equation}
\begin{dcases}
     \delta \mos=-\frac{1}{2}M_S^2 +4 \kappa e^2 M_V^2,\\*
   \los+ \delta \los =\frac{e^2}{2}\frac{M_S^2}{M_V^2}-4 \kappa e^4 \left(\log \frac{M_V^2}{\mu^2}+\frac{2}{3} + \eta\right).\label{eq:OS-counterterms}
\end{dcases}
\end{equation}
Introducing the ratio of the physical masses
\begin{equation}
    r=\frac{M_S^2}{M_V^2}
\end{equation}
and using the CW mass ratio of eq.~~\eqref{eq:mass-ratio} we obtain the following expression for the renormalised one-loop effective potential
\begin{equation}
    V(\f)=\kappa e^4 \f^4\left(\log\frac{e^2\f^2}{M_V^2}-\frac{3}{2} + \frac{r}{\rcw}\right) + 2 \kappa e^2 M_V^2\f^2\left(1-\frac{r}{\rcw}\right).\label{eq:OS-potential}
\end{equation}
We cannot further simplify this potential using the stationarity condition, as was done in the $\ms$ and CW cases, since we have already used this condition for fixing the counterterms. 

 The first observation is that the ratio of the physical masses $r$ turns out to be the key input parameter determining the properties of the potential, in contrast to the \textit{prediction} for the mass ratio in the $\ms$/CW schemes. Second, in the OS procedure, a nonvanishing (finite) mass counterterm is generated, which explicitly breaks the classical scale invariance. Since the scale symmetry in the analysed model is imposed only at classical level, there is no fundamental reason why it should be preserved once quantum corrections are taken into account. In fact, the CW and $\ms$ schemes are tailored to preserve the explicit scale invariance at loop level. The former explicitly demands that the renormalised mass term is zero via one of the renormalisation conditions. The other is based on the property of dimensional regularisation, which prevents the appearance of quadratic divergences. This is also the reason why in the $\ms$ and CW schemes one less parameter was needed to fully specify the model -- the mass term was ``invisible'' since it was set to zero.

\begin{figure}[!ht]
    \centering
    \includegraphics[width=.9\columnwidth]{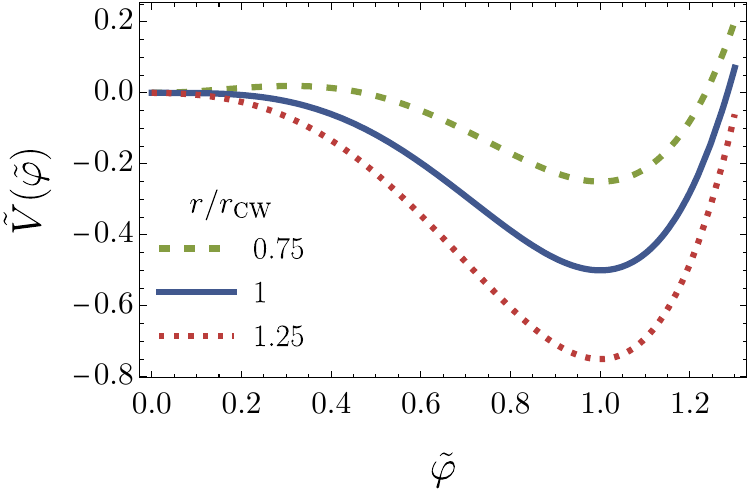}
    \caption{The rescaled potential of eq.~\eqref{eq:rescaled-potential} for $r/\rcw=0.75$ (long-dashed, green), $r/\rcw=1$ (solid, blue), $r/\rcw=1.25$ (short-dashed, red).}
    \label{fig:potential}
\end{figure}

We can now study the structure of the renormalised potential depending on the value of $r/\rcw$. This is illustrated in figure~\ref{fig:potential}, where we show the potential normalised by the mass scale set by $M_V$
\begin{equation}
    \tilde{V}(\tilde{\f})=\frac{V(\f)}{\kappa M_V^4}=\tilde{\f}^4\left(\log\tilde{\f}^2-\frac{3}{2} + \frac{r}{\rcw}\right) + 2 \tilde{\f}^2\left(1-\frac{r}{\rcw}\right),\label{eq:rescaled-potential}
\end{equation}
where $\tilde{\f}$ is a dimensionless field $\tilde{\f}=e \f/M_V$.
In this parameterisation, the shape of the potential only depends on the ratio $r/\rcw$, while changing the values of $M_V$ and $e$ does not change the picture.

If $r=\rcw$ (solid blue line), the mass term vanishes at one-loop level.  The potential of eq.~\eqref{eq:OS-potential} exactly coincides with the $\ms$ and CW result and has the familiar logarithmic shape, very flat around the origin, dominated by the term quartic in the field. When $r/\rcw<1$ (long-dashed green curve), the mass term in the potential is positive and it induces a minimum at the field-space origin and a barrier separating the symmetric and the broken (radiatively generated) minima. In this regime, symmetry breaking is even harder to achieve than in the standard RSB scenario, as the loop corrections have to cancel the tree-level quadratic and quartic terms. For $r/\rcw<0.5$ the symmetry-breaking minimum becomes only a local minimum of the potential. When $r/\rcw>1$ (red dotted curve), the mass term is negative and contributes to symmetry breaking as in the classical version of the Brout--Englert--Higgs mechanism. If $r/\rcw$ is close to one, it is only a small correction to the RSB mechanism but if it dominates over the logarithmic term, the potential has a parabolic shape around the origin. For $r/\rcw>1.5$, the sign of the $\tilde{\f}^4$ term becomes positive, contrary to what is observed in the common RSB analyses.

\section{Relating on-shell and \texorpdfstring{$\ms$}{msbar} results\label{sec:MS-OS}}
To further explore the relation between the $\ms$ and the OS schemes we will now relate the potential parameters defined in both schemes. To do that, we use the fact that the bare couplings are scheme-independent. We thus have
\begin{equation}
    \lambda_{\mathrm{bare}}=\lms+\delta\lms=\los+\delta\los.\label{eq:relating-OS-MS}
\end{equation}
From this relation, we obtain $\lms$ in terms of the physical masses
\begin{equation}
    \lms=\los+\delta\los-\delta\lms=\frac{e^2}{4}\frac{M_S^2}{M_V^2}-4\kappa e^4 \left(\log\frac{M_V^2}{\mu^2}+\frac{2}{3}\right),\label{eq:lambda-ms}
\end{equation}
where the expressions for the counterterms from eqs.~\eqref{eq:MS-counterterms} and \eqref{eq:OS-counterterms} have been used. 
The same reasoning can be employed to rewrite the $\ms$ mass parameter in terms of the physical quantities. Using the fact that $\mos=\delta\mms=0$ we get
\begin{equation}
    \mms=\delta \mos.\label{eq:mass-ms}
\end{equation}
If we now substitute $\mms$ and $\lms$ from eqs.~\eqref{eq:mass-ms} and~\eqref{eq:lambda-ms} to the $\ms$-renormalised potential of eq.~\eqref{eq:MS-potential}, we recover the expression for the OS-renormalised potential of eq.~\eqref{eq:OS-potential}, as expected. The relation of eq.~\eqref{eq:mass-ms}, however, tells us that if we want to capture the same physics that is described by the model in the OS, in the $\ms$ scheme, a mass term appears. That is not visible from inside the $\ms$ scheme because in dimensional regularisation no quadratic divergences appear. Also, no direct link to the physical masses is made during the procedure of $\ms$ renormalisation so there is no need to introduce a mass counterterm.

\begin{figure}[!ht]
    \centering
    \includegraphics[width=.9\columnwidth]{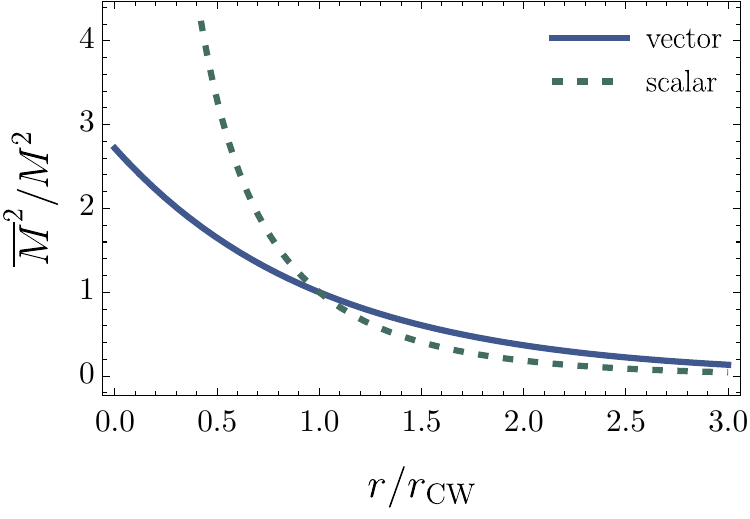}
    \caption{The ratio of the running mass to the physical one for the vector (solid line) and the scalar (dashed line).}
    \label{fig:masses}
\end{figure}

It is likely more insightful to explore the relation between the OS and $\ms$ masses. This can be achieved by equating the expression for $\lms$ from eq.~\eqref{eq:lambda-ms} with the one obtained from the minimisation condition in the $\ms$ scheme, eq.~\eqref{eq:lambda-e-MS}, using the fact that $e$ needs not to be renormalised and hence it is the same in both schemes. This leads to the following  relation between the OS and $\ms$ vector masses
\begin{equation}
    \frac{\overline{M}_V^2}{M_V^2}=\exp\left(1-\frac{r}{\rcw}\right).
\end{equation}
Using eq.~\eqref{eq:mass-ratio} we obtain a similar result for the scalar mass
\begin{equation}
        \frac{\overline{M}_S^2}{M_S^2}=\frac{\rcw}{r}\exp\left(1-\frac{r}{\rcw}\right).
\end{equation}
The behaviour of these ratios is illustrated in figure~\ref{fig:masses} (solid line for the vector and dashed line for the scalar). 
If the physical mass ratio coincides with the CW value, the running masses exactly coincide with the physical ones (in the zero-momentum approximation) and the $\ms$ and CW descriptions work perfectly. On the other hand, if the mass ratio is different but we try to use a model that has no mass term, our predictions will quickly deviate from the physical values. In particular, the OS scalar mass to the running mass ratio diverges as $r\to 0$. This is due to the disappearance of the global symmetry-breaking minimum in this limit (see figure~\ref{fig:potential}). For $r/\rcw\gg 1$, the OS to running mass ratios tend to zero. In this case, our description should contain a mass term already at classical level.

\section{Loop-generated vs tree-level mass term\label{sec:mass-terms}}

One may ask the following question -- if a mass term is generated in the renormalisation procedure, could we, equally well, include it in the potential from the very beginning? To understand that, we can conduct the same analysis starting from the potential of eq.~\eqref{eq:V-tree} supplemented with a mass term
\begin{equation}
    \Vtree(\f)=\frac{1}{4}\lambda \f^4 + \frac{1}{2}m^2 \f^2.
\end{equation}
We can apply the same renormalisation conditions as before, see eq.~\eqref{eq:on-shell-Veff}. Then it is only possible to determine the combinations $\lambda+\delta \lambda$, $m^2+\delta m^2$. The result is the same as in the massless case given in eq.~\eqref{eq:OS-counterterms}, with $m^2$ added on the left-hand side of the first equation. Hence, it does not matter whether a nonzero mass term had been included at tree level, if we impose the on-shell renormalisation conditions on the full one-loop potential.

Having a (negative) tree-level mass term that allows for symmetry breaking already at tree level, we can relate the physical masses, $M_S$ and $M_V$, to Lagrangian parameters already at tree level, using the first and the second derivative of the classical potential. Applying the relation $\langle \f \rangle =M_V/e$ we have
\begin{equation}
\begin{dcases}
   m^2 =-\frac{1}{2}M_S^2, \\*
  \lambda  = \frac{e^2}{2}\frac{M_S^2}{M_V^2}. \label{eq:tree-level-with-m2}
\end{dcases}
\end{equation}
Now we can renormalise the potential using the more standard conditions listed in eq.~\eqref{eq:on-shell-Vtree}, which state that the one-loop corrections should not alter the tree-level values of the scalar mass and VEV. This yields
\begin{equation}
\begin{dcases}
   \delta m^2=4 \kappa e^2 {M_V^2}, \\*
  \delta \lambda = -4 \kappa e^4 \left(\log \frac{M_V^2}{\mu^2}+\eta + \frac{2}{3}\right).\label{eq:counterterms-with-m2-2}
\end{dcases}
\end{equation}
The relations of eqs.~\eqref{eq:counterterms-with-m2-2} and~\eqref{eq:tree-level-with-m2} correspond to a split of eq.~\eqref{eq:OS-counterterms} (supplemented with the $m^2$ term as described above) into tree-level and one-loop parts. This approach allows us to relate the tree-level Lagrangian parameters with the physical parameters but the final potential, expressed in terms of the physical parameters, will look the same.

The discussion of this subsection shows that the split between the tree-level and loop-generated mass term is to some extent artificial. Nonetheless, to describe a model of scalar electrodynamics with arbitrary masses for the scalar and the vector, we need a potential with a term quadratic in the field. If the ratio of the masses is very special and equal to $\rcw$ as ``predicted'' in the $\ms$ or CW scheme (see eq.~\eqref{eq:mass-ratio}), one can describe it with a potential that is classically scale invariant. In this case, at loop level no mass term is generated and the scale symmetry is only broken by logarithmic effects. On the other hand, in the very same case of the CW mass ratio, one can start at tree level with a description that is not scale invariant and contains a mass term as in eq.~\eqref{eq:tree-level-with-m2}. Then, the loop corrections will cancel this mass term and in the one-loop potential the symmetry will only be broken by logarithmic effects -- an effective ``radiative (explicit) symmetry restoration''. This shows that a logarithmic one-loop potential can be attained if tree-level and one-loop terms are of the same order (this condition is expressed by $r=\rcw$), irrespective of whether there is a mass term in the tree-level potential.

Of course, if the mass ratio is far away from $\rcw$, the last approach to renormalisation of the effective potential presented above is the most suitable. In such a case, the tree-level term dominates, and the one-loop contribution gives only a small correction to the leading-order results. Then there is a natural split between the leading tree-level order and the subleading one-loop order. The perturbative expansion should follow the loop expansion.

This leads us to the conclusion that in the parameter space of scalar QED, there is a subspace, parameterised by $r=\rcw$, where tree-level and one-loop terms conspire to produce a logarithmic potential at one-loop level. In this subclass, one should use the scaling of couplings as a guideline to conduct perturbative expansions, instead of the loop counting. Is it fair to say that this subclass corresponds to classical scale invariance? Not exactly, since, as we have seen the same result is obtained even with a tree-level mass term, if the ratio of physical masses is equal to $\rcw$. It is, nonetheless, true that this case corresponds to radiative symmetry breaking, as the one-loop and tree-level contributions conspire to produce a potential with a non-trivial minimum, generated by the loop-induced logarithm, not by a term quadratic in the field. Therefore, the condition $r=\rcw$ can be considered as an on-shell condition for the realisation of RSB.

One remark is in order. In the whole reasoning above, we have neglected the scalar contribution to the effective potential. This is well justified in the case of RSB when $\lambda$ scales as $e^4$. Then the scalar contribution scales as $\lambda^2 \sim e^8$ and is a higher-order effect. This should hold as long as the mass ratio is close to the CW one. If we drift far away from it, the scalar effects (and the gauge-dependent terms discussed in section~\ref{sec:model}) may become more important. They are, however, usually expected to be subleading compared to the vector ones.

\section{Summary and discussion\label{sec:summary}}

In this article, we have studied an archetypal model of radiative symmetry breaking -- massless scalar QED -- from the on-shell perspective. We parameterised the model with the physical masses of the vector and scalar particles, and the electric charge. The renormalisation conditions were chosen such that the scalar mass corresponds to the pole of the propagator (in the zero-momentum approximation) and the vector mass is defined via the tree-level relation at the one-loop minimum. We have compared the obtained results with the ones from the $\ms$ and CW schemes. Below we summarise the main findings:
\begin{itemize}
    \item[--] If we choose arbitrary masses for the vector and the scalar, then at one-loop level a mass (counter)term is generated. This mass term is controlled by the ratio of the physical masses. If the ratio coincides with the value predicted in $\ms$ and CW schemes, $\rcw$, the mass term vanishes.
    \item[--] For a given ratio of masses we can equally well use a potential containing a mass term. If the ratio of masses is equal to $\rcw$, in the renormalised potential the mass term is cancelled by a counterterm and the resulting potential has the logarithmic form, characteristic of scale-invariant models.
    \item[--] In general, the scalar-to-vector mass ratio controls the properties of the renormalised potential. For $r<\rcw$, a positive mass term is present, which impedes symmetry breaking. For $r<0.5\rcw$, the symmetry-breaking minimum becomes a local one. For $r>\rcw$, there is a negative mass term that facilitates symmetry breaking. For $r$ deviating slightly from $\rcw$ the quadratic term is only a small correction to the logarithmic potential, but for larger $r$, the potential becomes polynomial as in the usual Brout--Englert--Higgs mechanism.
    \item[--] If $r$ is close to $\rcw$ the running masses approximate the physical masses well. The approximation becomes exponentially worse with $r$ deviating from $\rcw$.
    \item[--] Models with logarithmic potentials are not necessarily classically scale symmetric. They are rather a subclass of more general massive models, in which tree-level and one-loop terms conspire to cancel mass terms. In this sense, the term radiative symmetry breaking is more general than classical scale symmetry. In the case of scalar QED the sub-model with RSB is defined by $r=\rcw$. Therefore, the condition $r=\rcw$ should not be considered a prediction of massless scalar QED~\cite{Coleman:1973jx} but rather an assumption necessary for it to be realised, an on-shell version of the typical condition necessary for radiative symmetry breaking, $\lambda \sim e^4$. 
\end{itemize}

These observations can contribute to the discussion on whether classical scale symmetry is a solution to the hierarchy problem, dating back to the paper by Bardeen~\cite{Bardeen:1995}. The common argument is simple -- with classical scale symmetry and dimensional regularisation, no quadratic divergence is generated so the hierarchy problem of the SM is cured. However, as we show in the present paper, to attain a potential with vanishing renormalised mass term, one needs to tune the physical masses to give the CW ratio, $\rcw$. Therefore, the tuning of the Higgs (scalar) mass is traded for tuning the scalar-to-vector mass ratio.

The main motivation for performing this exercise in renormalisation was to verify whether the extremely promising predictions for GW signals from supercooled phase transitions in classically scale-invariant models are robust to the renormalisation scheme change. The appearance of a strong GW signal relies on the absence of the mass term in the renormalised potential, thus, generating such a term in the OS scheme could change the predictions dramatically. A vast majority (if not all) of works on classically scale-invariant models utilise the $\ms$ (or CW) scheme, we are not aware of a reference that would use the OS scheme. This paper aims to fill this gap.

Our findings are reassuring. If one assumes $r=\rcw$, no mass term is generated in the renormalised potential and the results of $\ms$ and CW schemes coincide with the OS ones. One should be aware, however, that $r=\rcw$ is rather an assumption in the procedure than a prediction of the model. This means that to find a logarithmic potential suitable for supercooling one needs to tune the physical masses such that their ratio is close to $\rcw$.\footnote{Of course, in more complicated models the criterion of the scalar-to-vector mass ratio has to be substituted with an adequate analogue.} Having that, one can proceed with $\ms$ or OS schemes, and the results should be the same. In particular, the predictions of strong GW signals originating from supercooled phase transitions obtained within the $\ms$ scheme hold and such scenarios are prime candidates for being observed with LISA. 

If the mass ratio does not coincide with $\rcw$, a mass term should be included in the potential. One may ask -- will this dramatically change the predictions for the GW signature of the phase transition? Ref.~\cite{Levi:2022bzt} presented a detailed study of models where scale symmetry is softly broken  (by a scalar mass term or by a cubic coupling). It was shown there that the inclusion of dimensionful couplings in the tree-level Lagrangian influences the range of the gauge coupling of the new gauge group for which supercooling is present. In particular, negative mass term or cubic coupling allows for supercooling in a broader range of the gauge coupling (although the signal can be weaker, yet still observable), while positive dimensionful couplings act in the opposite direction. For a fixed value of the new gauge coupling an inclusion of a very small mass term only affects the nucleation temperature and thus the GW signal remains unchanged. For larger mass terms the phase transition becomes faster than in the scale-invariant case, which makes the signal weaker. This effect is more visible for the values of the gauge coupling that in a pure scale-invariant scenario would lead to the strongest supercooling. Overall, the analysis of Ref.~\cite{Levi:2022bzt} shows that even if the masses are not tuned to exactly give the $\rcw$ ratio, supercooling can be expected. This is in accord with the results presented here. Even if $r\neq \rcw$ we can insist on parameterising the model by a classically scale-invariant potential. Then, a mass counterterm will appear in the OS scheme, however, the same potential renormalised with the $\ms$ scheme will appear to have a logarithmic form. Such a potential yields a strong GW signal. Moreover, its predictions for the masses of particles will be approximately correct as long as $r$ is close to $\rcw$.

To sum up, if the physical masses are not exactly tuned to the Coleman-Weinberg ratio, a mass term in the potential is generated. Nonetheless, it does not destroy the conclusions regarding the production of a strong GW signal during a supercooled phase transition, as long as the deviation from the exact Coleman-Weinberg limit is small.

\section*{Acknowledgements}
We thank  Gabriel Lourenço and Rui Santos  for numerous fruitful discussions and for reading the ma\-nu\-script. We are grateful to Miko{\l}aj Misiak for an interesting discussion. This research was funded by the National Science Centre, Poland, through the SONATA project number 2018/31/D/ST2/03302 and the OPUS project number 2023/49/B/ST2/02782. For the purpose of Open Access, the authors have applied a CC-BY public copyright licence to any Author Accepted Manuscript version arising from this submission.

\bibliographystyle{elsarticle-num} 
\bibliography{conformal-reno}

\begin{thebibliography}{10}
\expandafter\ifx\csname url\endcsname\relax
  \def\url#1{\texttt{#1}}\fi
\expandafter\ifx\csname urlprefix\endcsname\relax\def\urlprefix{URL }\fi
\expandafter\ifx\csname href\endcsname\relax
  \def\href#1#2{#2} \def\path#1{#1}\fi

\bibitem{LISACosmologyWorkingGroup:2022jok}
P.~Auclair, et~al., {Cosmology with the Laser Interferometer Space Antenna},
  Living Rev. Rel. 26~(1) (2023) 5.
\newblock \href {http://arxiv.org/abs/2204.05434} {\path{arXiv:2204.05434}},
  \href {https://doi.org/10.1007/s41114-023-00045-2}
  {\path{doi:10.1007/s41114-023-00045-2}}.

\bibitem{Caprini:2015zlo}
C.~Caprini, et~al., {Science with the space-based interferometer eLISA. II:
  Gravitational waves from cosmological phase transitions}, JCAP 04 (2016) 001.
\newblock \href {http://arxiv.org/abs/1512.06239} {\path{arXiv:1512.06239}},
  \href {https://doi.org/10.1088/1475-7516/2016/04/001}
  {\path{doi:10.1088/1475-7516/2016/04/001}}.

\bibitem{Caprini:2019egz}
C.~Caprini, et~al., {Detecting gravitational waves from cosmological phase
  transitions with LISA: an update}, JCAP 03 (2020) 024.
\newblock \href {http://arxiv.org/abs/1910.13125} {\path{arXiv:1910.13125}},
  \href {https://doi.org/10.1088/1475-7516/2020/03/024}
  {\path{doi:10.1088/1475-7516/2020/03/024}}.

\bibitem{Kajantie:1996mn}
K.~Kajantie, M.~Laine, K.~Rummukainen, M.~E. Shaposhnikov, {Is there a~ hot
  electroweak phase transition at $m_H \gtrsim m_W$?}, Phys. Rev. Lett. 77
  (1996) 2887--2890.
\newblock \href {http://arxiv.org/abs/hep-ph/9605288}
  {\path{arXiv:hep-ph/9605288}}, \href
  {https://doi.org/10.1103/PhysRevLett.77.2887}
  {\path{doi:10.1103/PhysRevLett.77.2887}}.

\bibitem{Gurtler:1997hr}
M.~Gurtler, E.-M. Ilgenfritz, A.~Schiller, {Where the electroweak phase
  transition ends}, Phys. Rev. D 56 (1997) 3888--3895.
\newblock \href {http://arxiv.org/abs/hep-lat/9704013}
  {\path{arXiv:hep-lat/9704013}}, \href
  {https://doi.org/10.1103/PhysRevD.56.3888}
  {\path{doi:10.1103/PhysRevD.56.3888}}.

\bibitem{Csikor:1998eu}
F.~Csikor, Z.~Fodor, J.~Heitger, {Endpoint of the hot electroweak phase
  transition}, Phys. Rev. Lett. 82 (1999) 21--24.
\newblock \href {http://arxiv.org/abs/hep-ph/9809291}
  {\path{arXiv:hep-ph/9809291}}, \href
  {https://doi.org/10.1103/PhysRevLett.82.21}
  {\path{doi:10.1103/PhysRevLett.82.21}}.

\bibitem{Randall:2006}
L.~Randall, G.~Servant, {Gravitational waves from warped spacetime}, JHEP 05
  (2007) 054.
\newblock \href {http://arxiv.org/abs/hep-ph/0607158}
  {\path{arXiv:hep-ph/0607158}}, \href
  {https://doi.org/10.1088/1126-6708/2007/05/054}
  {\path{doi:10.1088/1126-6708/2007/05/054}}.

\bibitem{Konstandin:2010}
T.~Konstandin, G.~Nardini, M.~Quiros, {Gravitational Backreaction Effects on
  the Holographic Phase Transition}, Phys. Rev. D82 (2010) 083513.
\newblock \href {http://arxiv.org/abs/1007.1468} {\path{arXiv:1007.1468}},
  \href {https://doi.org/10.1103/PhysRevD.82.083513}
  {\path{doi:10.1103/PhysRevD.82.083513}}.

\bibitem{Konstandin:2011}
T.~Konstandin, G.~Servant, {Cosmological Consequences of Nearly Conformal
  Dynamics at the TeV scale}, JCAP 1112 (2011) 009.
\newblock \href {http://arxiv.org/abs/1104.4791} {\path{arXiv:1104.4791}},
  \href {https://doi.org/10.1088/1475-7516/2011/12/009}
  {\path{doi:10.1088/1475-7516/2011/12/009}}.

\bibitem{Hambye:2013}
T.~Hambye, A.~Strumia, {Dynamical generation of the weak and Dark Matter
  scale}, Phys. Rev. D88 (2013) 055022.
\newblock \href {http://arxiv.org/abs/1306.2329} {\path{arXiv:1306.2329}},
  \href {https://doi.org/10.1103/PhysRevD.88.055022}
  {\path{doi:10.1103/PhysRevD.88.055022}}.

\bibitem{Jaeckel:2016}
J.~Jaeckel, V.~V. Khoze, M.~Spannowsky, {Hearing the signal of dark sectors
  with gravitational wave detectors}, Phys. Rev. D94~(10) (2016) 103519.
\newblock \href {http://arxiv.org/abs/1602.03901} {\path{arXiv:1602.03901}},
  \href {https://doi.org/10.1103/PhysRevD.94.103519}
  {\path{doi:10.1103/PhysRevD.94.103519}}.

\bibitem{Jinno:2016}
R.~Jinno, M.~Takimoto, {Probing a classically conformal B-L model with
  gravitational waves}, Phys. Rev. D95~(1) (2017) 015020.
\newblock \href {http://arxiv.org/abs/1604.05035} {\path{arXiv:1604.05035}},
  \href {https://doi.org/10.1103/PhysRevD.95.015020}
  {\path{doi:10.1103/PhysRevD.95.015020}}.

\bibitem{Marzola:2017}
L.~Marzola, A.~Racioppi, V.~Vaskonen, {Phase transition and gravitational wave
  phenomenology of scalar conformal extensions of the Standard Model}, Eur.
  Phys. J. C77~(7) (2017) 484.
\newblock \href {http://arxiv.org/abs/1704.01034} {\path{arXiv:1704.01034}},
  \href {https://doi.org/10.1140/epjc/s10052-017-4996-1}
  {\path{doi:10.1140/epjc/s10052-017-4996-1}}.

\bibitem{Hashino:2018}
K.~Hashino, M.~Kakizaki, S.~Kanemura, P.~Ko, T.~Matsui, {Gravitational waves
  from first order electroweak phase transition in models with the U(1)$_{X}$
  gauge symmetry}, JHEP 06 (2018) 088.
\newblock \href {http://arxiv.org/abs/1802.02947} {\path{arXiv:1802.02947}},
  \href {https://doi.org/10.1007/JHEP06(2018)088}
  {\path{doi:10.1007/JHEP06(2018)088}}.

\bibitem{Baldes:2018}
I.~Baldes, C.~Garcia-Cely, {Strong gravitational radiation from a simple dark
  matter model}, JHEP 05 (2019) 190.
\newblock \href {http://arxiv.org/abs/1809.01198} {\path{arXiv:1809.01198}},
  \href {https://doi.org/10.1007/JHEP05(2019)190}
  {\path{doi:10.1007/JHEP05(2019)190}}.

\bibitem{Prokopec:2018}
T.~Prokopec, J.~Rezacek, B.~\'Swie\.zewska, {Gravitational waves from conformal
  symmetry breaking}, JCAP 02 (2019) 009.
\newblock \href {http://arxiv.org/abs/1809.11129} {\path{arXiv:1809.11129}},
  \href {https://doi.org/10.1088/1475-7516/2019/02/009}
  {\path{doi:10.1088/1475-7516/2019/02/009}}.

\bibitem{Marzo:2018}
C.~Marzo, L.~Marzola, V.~Vaskonen, {Phase transition and vacuum stability in
  the classically conformal B\textendash{}L model}, Eur. Phys. J. C 79~(7)
  (2019) 601.
\newblock \href {http://arxiv.org/abs/1811.11169} {\path{arXiv:1811.11169}},
  \href {https://doi.org/10.1140/epjc/s10052-019-7076-x}
  {\path{doi:10.1140/epjc/s10052-019-7076-x}}.

\bibitem{Mohamadnejad:2019vzg}
A.~Mohamadnejad, {Gravitational waves from scale-invariant vector dark matter
  model: Probing below the neutrino-floor}, Eur. Phys. J. C 80~(3) (2020) 197.
\newblock \href {http://arxiv.org/abs/1907.08899} {\path{arXiv:1907.08899}},
  \href {https://doi.org/10.1140/epjc/s10052-020-7756-6}
  {\path{doi:10.1140/epjc/s10052-020-7756-6}}.

\bibitem{Kang:2020jeg}
Z.~Kang, J.~Zhu, {Scale-genesis by Dark Matter and Its Gravitational Wave
  Signal}, Phys. Rev. D 102~(5) (2020) 053011.
\newblock \href {http://arxiv.org/abs/2003.02465} {\path{arXiv:2003.02465}},
  \href {https://doi.org/10.1103/PhysRevD.102.053011}
  {\path{doi:10.1103/PhysRevD.102.053011}}.

\bibitem{Ellis:2020}
J.~Ellis, M.~Lewicki, V.~Vaskonen, {Updated predictions for gravitational waves
  produced in a strongly supercooled phase transition}, JCAP 11 (2020) 020.
\newblock \href {http://arxiv.org/abs/2007.15586} {\path{arXiv:2007.15586}},
  \href {https://doi.org/10.1088/1475-7516/2020/11/020}
  {\path{doi:10.1088/1475-7516/2020/11/020}}.

\bibitem{Azatov:2020nbe}
A.~Azatov, M.~Vanvlasselaer, {Phase transitions in perturbative walking
  dynamics}, JHEP 09 (2020) 085.
\newblock \href {http://arxiv.org/abs/2003.10265} {\path{arXiv:2003.10265}},
  \href {https://doi.org/10.1007/JHEP09(2020)085}
  {\path{doi:10.1007/JHEP09(2020)085}}.

\bibitem{Mohamadnejad:2021tke}
A.~Mohamadnejad, {Electroweak phase transition and gravitational waves in a
  two-component dark matter model}, JHEP 03 (2022) 188.
\newblock \href {http://arxiv.org/abs/2111.04342} {\path{arXiv:2111.04342}},
  \href {https://doi.org/10.1007/JHEP03(2022)188}
  {\path{doi:10.1007/JHEP03(2022)188}}.

\bibitem{Dasgupta:2022isg}
A.~Dasgupta, P.~S.~B. Dev, A.~Ghoshal, A.~Mazumdar, {Gravitational wave pathway
  to testable leptogenesis}, Phys. Rev. D 106~(7) (2022) 075027.
\newblock \href {http://arxiv.org/abs/2206.07032} {\path{arXiv:2206.07032}},
  \href {https://doi.org/10.1103/PhysRevD.106.075027}
  {\path{doi:10.1103/PhysRevD.106.075027}}.

\bibitem{Kierkla:2022odc}
M.~Kierkla, A.~Karam, B.~{\'S}wie{\.z}ewska, {Conformal model for gravitational
  waves and dark matter: a status update}, JHEP 03 (2023) 007.
\newblock \href {http://arxiv.org/abs/2210.07075} {\path{arXiv:2210.07075}},
  \href {https://doi.org/10.1007/JHEP03(2023)007}
  {\path{doi:10.1007/JHEP03(2023)007}}.

\bibitem{Frandsen:2022klh}
M.~T. Frandsen, M.~Heikinheimo, M.~E. Thing, K.~Tuominen, M.~Rosenlyst, {Vector
  dark matter in supercooled Higgs portal models}, Phys. Rev. D 108~(1) (2023)
  015033.
\newblock \href {http://arxiv.org/abs/2301.00041} {\path{arXiv:2301.00041}},
  \href {https://doi.org/10.1103/PhysRevD.108.015033}
  {\path{doi:10.1103/PhysRevD.108.015033}}.

\bibitem{Kierkla:2023von}
M.~Kierkla, B.~{\'S}wie{\.z}ewska, T.~V.~I. Tenkanen, J.~van~de Vis,
  {Gravitational waves from supercooled phase transitions: dimensional
  transmutation meets dimensional reduction}, JHEP 02 (2024) 234.
\newblock \href {http://arxiv.org/abs/2312.12413} {\path{arXiv:2312.12413}},
  \href {https://doi.org/10.1007/JHEP02(2024)234}
  {\path{doi:10.1007/JHEP02(2024)234}}.

\bibitem{Chun:2023ezg}
E.~J. Chun, T.~P. Dutka, T.~H. Jung, X.~Nagels, M.~Vanvlasselaer,
  {Bubble-assisted leptogenesis}, JHEP 09 (2023) 164.
\newblock \href {http://arxiv.org/abs/2305.10759} {\path{arXiv:2305.10759}},
  \href {https://doi.org/10.1007/JHEP09(2023)164}
  {\path{doi:10.1007/JHEP09(2023)164}}.

\bibitem{Ratzinger:2020koh}
W.~Ratzinger, P.~Schwaller, {Whispers from the dark side: Confronting light new
  physics with NANOGrav data}, SciPost Phys. 10~(2) (2021) 047.
\newblock \href {http://arxiv.org/abs/2009.11875} {\path{arXiv:2009.11875}},
  \href {https://doi.org/10.21468/SciPostPhys.10.2.047}
  {\path{doi:10.21468/SciPostPhys.10.2.047}}.

\bibitem{NANOGrav:2021flc}
Z.~Arzoumanian, et~al., {Searching for Gravitational Waves from Cosmological
  Phase Transitions with the NANOGrav 12.5-Year Dataset}, Phys. Rev. Lett.
  127~(25) (2021) 251302.
\newblock \href {http://arxiv.org/abs/2104.13930} {\path{arXiv:2104.13930}},
  \href {https://doi.org/10.1103/PhysRevLett.127.251302}
  {\path{doi:10.1103/PhysRevLett.127.251302}}.

\bibitem{Gouttenoire:2023bqy}
Y.~Gouttenoire, {First-Order Phase Transition Interpretation of Pulsar Timing
  Array Signal Is Consistent with Solar-Mass Black Holes}, Phys. Rev. Lett.
  131~(17) (2023) 171404.
\newblock \href {http://arxiv.org/abs/2307.04239} {\path{arXiv:2307.04239}},
  \href {https://doi.org/10.1103/PhysRevLett.131.171404}
  {\path{doi:10.1103/PhysRevLett.131.171404}}.

\bibitem{Madge:2023dxc}
E.~Madge, E.~Morgante, C.~Puchades-Ib\'a\~nez, N.~Ramberg, W.~Ratzinger,
  S.~Schenk, P.~Schwaller, {Primordial gravitational waves in the nano-Hertz
  regime and PTA data \textemdash{} towards solving the GW inverse problem},
  JHEP 10 (2023) 171.
\newblock \href {http://arxiv.org/abs/2306.14856} {\path{arXiv:2306.14856}},
  \href {https://doi.org/10.1007/JHEP10(2023)171}
  {\path{doi:10.1007/JHEP10(2023)171}}.

\bibitem{Figueroa:2023zhu}
D.~G. Figueroa, M.~Pieroni, A.~Ricciardone, P.~Simakachorn, {Cosmological
  Background Interpretation of Pulsar Timing Array Data}, Phys. Rev. Lett.
  132~(17) (2024) 171002.
\newblock \href {http://arxiv.org/abs/2307.02399} {\path{arXiv:2307.02399}},
  \href {https://doi.org/10.1103/PhysRevLett.132.171002}
  {\path{doi:10.1103/PhysRevLett.132.171002}}.

\bibitem{Ellis:2023oxs}
J.~Ellis, M.~Fairbairn, G.~Franciolini, G.~H\"utsi, A.~Iovino, M.~Lewicki,
  M.~Raidal, J.~Urrutia, V.~Vaskonen, H.~Veerm\"ae, {What is the source of the
  PTA GW signal?}, Phys. Rev. D 109~(2) (2024) 023522.
\newblock \href {http://arxiv.org/abs/2308.08546} {\path{arXiv:2308.08546}},
  \href {https://doi.org/10.1103/PhysRevD.109.023522}
  {\path{doi:10.1103/PhysRevD.109.023522}}.

\bibitem{NANOGrav:2023gor}
G.~Agazie, et~al., {The NANOGrav 15 yr Data Set: Evidence for a
  Gravitational-wave Background}, Astrophys. J. Lett. 951~(1) (2023) L8.
\newblock \href {http://arxiv.org/abs/2306.16213} {\path{arXiv:2306.16213}},
  \href {https://doi.org/10.3847/2041-8213/acdac6}
  {\path{doi:10.3847/2041-8213/acdac6}}.

\bibitem{EPTA:2023fyk}
J.~Antoniadis, et~al., {The second data release from the European Pulsar Timing
  Array - III. Search for gravitational wave signals}, Astron. Astrophys. 678
  (2023) A50.
\newblock \href {http://arxiv.org/abs/2306.16214} {\path{arXiv:2306.16214}},
  \href {https://doi.org/10.1051/0004-6361/202346844}
  {\path{doi:10.1051/0004-6361/202346844}}.

\bibitem{Reardon:2023gzh}
D.~J. Reardon, et~al., {Search for an Isotropic Gravitational-wave Background
  with the Parkes Pulsar Timing Array}, Astrophys. J. Lett. 951~(1) (2023) L6.
\newblock \href {http://arxiv.org/abs/2306.16215} {\path{arXiv:2306.16215}},
  \href {https://doi.org/10.3847/2041-8213/acdd02}
  {\path{doi:10.3847/2041-8213/acdd02}}.

\bibitem{Bai:2021ibt}
Y.~Bai, M.~Korwar, {Cosmological constraints on first-order phase transitions},
  Phys. Rev. D 105~(9) (2022) 095015.
\newblock \href {http://arxiv.org/abs/2109.14765} {\path{arXiv:2109.14765}},
  \href {https://doi.org/10.1103/PhysRevD.105.095015}
  {\path{doi:10.1103/PhysRevD.105.095015}}.

\bibitem{Bringmann:2023opz}
T.~Bringmann, P.~F. Depta, T.~Konstandin, K.~Schmidt-Hoberg, C.~Tasillo, {Does
  NANOGrav observe a dark sector phase transition?}, JCAP 11 (2023) 053.
\newblock \href {http://arxiv.org/abs/2306.09411} {\path{arXiv:2306.09411}},
  \href {https://doi.org/10.1088/1475-7516/2023/11/053}
  {\path{doi:10.1088/1475-7516/2023/11/053}}.

\bibitem{Coleman:1973jx}
S.~R. Coleman, E.~J. Weinberg, {Radiative Corrections as the Origin of
  Spontaneous Symmetry Breaking}, Phys. Rev. D 7 (1973) 1888--1910.
\newblock \href {https://doi.org/10.1103/PhysRevD.7.1888}
  {\path{doi:10.1103/PhysRevD.7.1888}}.

\bibitem{Andreassen:2014eha}
A.~Andreassen, W.~Frost, M.~D. Schwartz, {Consistent Use of Effective
  Potentials}, Phys. Rev. D 91~(1) (2015) 016009.
\newblock \href {http://arxiv.org/abs/1408.0287} {\path{arXiv:1408.0287}},
  \href {https://doi.org/10.1103/PhysRevD.91.016009}
  {\path{doi:10.1103/PhysRevD.91.016009}}.

\bibitem{Espinosa:2016uaw}
J.~R. Espinosa, M.~Garny, T.~Konstandin, {Interplay of Infrared Divergences and
  Gauge-Dependence of the Effective Potential}, Phys. Rev. D 94~(5) (2016)
  055026.
\newblock \href {http://arxiv.org/abs/1607.08432} {\path{arXiv:1607.08432}},
  \href {https://doi.org/10.1103/PhysRevD.94.055026}
  {\path{doi:10.1103/PhysRevD.94.055026}}.

\bibitem{Andreassen:2013hpa}
A.~J. Andreassen, {Gauge Dependence of the Quantum Field Theory Effective
  Potential}, Master's thesis, Norwegian U. Sci. Tech. (2013).

\bibitem{Loebbert:2018xsd}
F.~Loebbert, J.~Miczajka, J.~Plefka, {Consistent Conformal Extensions of the
  Standard Model}, Phys. Rev. D 99~(1) (2019) 015026.
\newblock \href {http://arxiv.org/abs/1805.09727} {\path{arXiv:1805.09727}},
  \href {https://doi.org/10.1103/PhysRevD.99.015026}
  {\path{doi:10.1103/PhysRevD.99.015026}}.

\bibitem{Kang:1974yj}
J.~S. Kang, {Gauge Invariance of the Scalar-Vector Mass Ratio in the
  Coleman-Weinberg Model}, Phys. Rev. D 10 (1974) 3455.
\newblock \href {https://doi.org/10.1103/PhysRevD.10.3455}
  {\path{doi:10.1103/PhysRevD.10.3455}}.

\bibitem{Bardeen:1995}
W.~A. Bardeen, {On naturalness in the standard model}, in: {Ontake Summer
  Institute on Particle Physics Ontake Mountain, Japan, August 27-September 2,
  1995}, 1995.

\bibitem{Levi:2022bzt}
N.~Levi, T.~Opferkuch, D.~Redigolo, {The supercooling window at weak and strong
  coupling}, JHEP 02 (2023) 125.
\newblock \href {http://arxiv.org/abs/2212.08085} {\path{arXiv:2212.08085}},
  \href {https://doi.org/10.1007/JHEP02(2023)125}
  {\path{doi:10.1007/JHEP02(2023)125}}.

\end{thebibliography}

\end{document}